\documentclass[12pt]{article}
\pdfoutput=1

\usepackage{amsmath,amssymb,lmodern}
\usepackage{physics}
\usepackage{slashed}
\usepackage{authblk}
\usepackage{fancyhdr}
\usepackage{jheppub}

\newcommand{\nc}{\newcommand}
\nc{\La}{\Lambda}
\nc{\rnc}{\renewcommand }
\nc{\m}{\mu}
\nc{\f}{\varphi}
\nc{\n}{\nu}
\rnc{\k}{\kappa}
\rnc{\d}{\mathrm{d}}
\nc{\p}{\partial}
\rnc{\d}{\delta}
\nc{\g}{\gamma}
\rnc{\O}{\mathcal{O}}
\rnc{\t}{\tau}
\nc{\D}{\Delta}
\nc{\nn}{\nonumber}
\nc{\e}{\epsilon}
\nc{\w}{\omega}
\rnc{\a}{\alpha}
\rnc{\b}{\beta}
\rnc{\[}{\begin{equation}}
\rnc{\]}{\end{equation}}
\rnc{\(}{\left(}
\rnc{\)}{\right)}
\rnc{\l}{\lambda}
\rnc{\H}{\mathcal{H}}

\def\r{\rho}
\nc{\x}{\chi}
\rnc{\Lambda}{\kappa}
\nc{\<}{\left<}
\rnc{\>}{\right>}
\rnc{\|}{\Big\vert}

\title{Leading Order Anomalous Dimensions at the Wilson-Fisher Fixed Point from CFT}

\author[a]{Konstantinos Roumpedakis}

\affiliation[a]{C. N. Yang Institute for Theoretical Physics\\
Stony Brook University\\
Stony Brook, NY 11794, USA. }

\emailAdd{Konstantinos.roumpedakis@stonybrook.edu}

\abstract{In this paper we consider $\phi^4$ theory in $4-\e$ dimensions at the Wilson-Fisher fixed point where the theory becomes conformal. We extend the method in \cite{Slava} for calculating the leading order term in the anomalous dimensions of some operators with spin. This method involves mostly symmetry arguments and reduces the process for calculating anomalous dimensions to some Wick contractions in the corresponding free theory. We apply this method in the case of operators with two and three fields whose twist is equal to the number of fields they contain, and we rederive known results for their anomalous dimensions. We also calculate the leading term in the anomalous dimensions of operators with spin two and three. In addition, we find expressions for the primary operators of the free theory, for arbitrary spin and number of fields, whose twist remains equal to the number of fields.}

\preprint{YITP-SB-16-52}
\begin{document}

\maketitle

\numberwithin{equation}{section}

\section{Introduction}

	Recently there has been an enormous effort to calculate physical quantities in quantum field theory (QFT), relying mostly on the underlying symmetries of the theory and other fundamental principles like unitarity, to avoid the traditional tedious calculations of Feynman diagrams. So far, there have been several different approaches with different starting points, but all are based on the idea that symmetries alone can be used to extract a lot of information about the theory under consideration. The nature of these techniques suggests that they are more effective when there is an enhanced symmetry, such as conformal or supersymmetry. 
	
	This story goes back to the early seventies with the idea of the conformal bootstrap \cite{Polyakov,Belavin,Ferrara}. This idea drew a lot of attention during the last few years and was recently resurrected and applied to higher dimensional theories \cite{Rattazzi:2008pe}, leading to some strikingly precise numerical results \cite{Simmons-Duffin:2015qma,Kos:2016ysd}. In addition, recently there have been analytical approaches to the conformal bootstrap in conformal theories using the Mellin transformation \cite{Gopakumar:2016wkt,Gopakumar:2016cpb,Dey:2016mcs}, and others using large spin perturbation series \cite{Kaviraj:2015cxa,Kaviraj:2015xsa,Alday:2016njk,Alday:2016jfr,Aharony:2016dwx}. 

	In this work, we consider a technique, for studying a QFT with conformal symmetry (CFT), which first appeared in \cite{Slava}. This technique makes use of the fact that conformal symmetry fixes the three-point function of primary operators along with the idea of the operator product expansion (OPE). Initially this method was applied to $\phi^4$ theory in $4-\e$ dimensions, which has the non-trivial Wilson-Fisher fixed point, and was later applied to the critical scalar field theory in other dimensions \cite{Basu:2015gpa,Nii:2016lpa,Hasegawa:2016piv,Bashmakov:2016pcg} as well as to the Gross-Neveu model in $2+\e$ dimensions \cite{Ghosh:2015opa,Raju:2015fza}. A similar method, that is applicable only to a small subset of operators, namely the primary operators with two fields, had previously appeared in \cite{Anselmi:1998ms,Belitsky,Skvortsov:2015pea,Manashov:2015fha,Giombi}. For other techniques see also \cite{Sen:2015doa,Gliozzi:2016ysv}
	
	We consider only the $\phi^4$ theory in $4-\e$ dimensions. In this approach, $\e$ is a small number which serves as a perturbation parameter in the so called {\it $\e$-expansion}. The motivation behind this method is the well-known fact that $\phi^4$ theory describes the Ising model in three dimensions and that some of its critical exponents are related to the anomalous dimensions of some operators. In principle, one can compute a physical quantity as a power series in $\e$, use it to get a closed form answer, and then analytically continue to other values of $\e$ and in particular to $\e\rightarrow 1$. However, in this way one needs to know an infinite number of terms in the expansion and therefore we typically consider other methods for making estimates, e.g. the Pad\'e approximation \cite{Guida:1998bx,ZinnJustin:1999bf}. We should also note here that the $\e-$expansion is well known to be non-convergent, but it has been shown to be Borel summable \cite{Brezin:1976vw}. In this paper, we are interested in the method per se, leaving aside its physical applications.

	The goal of this article is to generalize the method in \cite{Slava} to a broader class of primary operators appearing in the theory, namely the operators whose twist (scale dimension minus spin) is equal to the number of fields they contain. In section \ref{sec:method}, after we have summarized the results of \cite{Slava}, we describe the generalization of this method  to include operators with spin, and we calculate the leading term in the anomalous dimensions of operators with spin two and three. This method reduces the calculations of anomalous dimensions to some free theory calculations of appropriate Green's functions of primary operators. In section \ref{sec:primaries} we construct the primary operators of the free theory in a suitable way for this method. In section \ref{sec:currents} we apply the method to operators with two fields, which are conserved currents in the free theory, and we rederive the well-known result for their anomalous dimensions.

	It is known in the literature \cite{Kehrein:1992fn,Braun:2003rp} that operators with more than two fields are degenerate for high enough spin and the method we describe in section \ref{sec:method} needs some modifications. In section \ref{sec:generalization}, we generalize the method to include  degenerate cases, and in section \ref{sec:three fields}, we rederive the known result for the anomalous dimensions of operators with three fields.

\section{The method\label{sec:method}}

We begin this section by reviewing the results of \cite{Slava}. We consider the $\phi^4$ theory in $d=4-\e$ dimensions with Lagrangian

\[L= \frac{1}{2} (\p \phi)^2+ \frac{1}{4!} g \m^\e \phi^4, \]
where $\m$ is a parameter with dimensions of mass, which in the usual approach of QFT plays the role of the renormalization scale. It is well-known that this theory has a non-trivial fixed point \cite{Wilson} where it becomes Conformal invariant. For small $\e$ this fixed point is perturbative and it is controlled by one-loop effects with coupling constant

\[g=g^*=\frac{16\pi^2}{3}\e +\mathcal{O}(\e^2). \label{fixed point}\]

Recently, a new method \cite{Slava} was invented for calculating the leading order correction for small $\e$ to the scaling dimensions of the operators in the theory, based only on the consistency of the conformal symmetry, without using the traditional diagrammatic way. 

From the CFT point of view, the theory is a collection of primary operators, each of them generating an irreducible representation (irrep) of the conformal algebra (\ref{algebra}). A single irrep consists of a primary operator and all its derivatives (descendants). An interaction in the CFT language is an extra constraint on the operator content of the theory. The effect of this constraint changes the dimensions of the operators (they acquire anomalous dimensions) and turns some primary operators to descendants (sort irreps of the CFT algebra become longer) leading to miltiplet recombination.

For simplicity, the field $\phi$ is rescaled as $\phi \rightarrow 2\pi \phi$ so that its two-point function is 

\[\<\phi(0) \phi(x)\>= \frac{1}{|x|^{2\D}},  \] 
where $\D$ is its scaling dimension. The equation of motion (because of the rescaling $\a= g/24 \pi^2$)

\[\Box \phi = \a \phi^3, \]
implies that the two irreps of the free theory generated by $\phi$ and $\phi^3$ combine to one single irrep. Using this fact, and the conformal invariance of the theory, the authors in \cite{Slava} showed that the position of the fixed point (\ref{fixed point}) can be determined for small $\e$ from the consistency of the theory. They also calculated the leading order in the anomalous dimension of the operators $\phi^n$ and they found\footnote{To be well-defined any composite operator is always assumed to be normal ordered}
\[\g_n=\frac{1}{6}n(n-1) \e. \]

The aim of this section is to generalize the method of \cite{Slava} for a broader class of operators with spin.

To begin with, the conformal algebra is

\begin{align}
&[D,P_\m]=iP_\m, \; [D,K_\m]=-iK_\m, \; [K_\m,P_\n]=-2iL_{\m\n}-2i\d_{\m\n}D \nn  \\
& [L_{\m\n},P_\r]=i(\d_{\m \r} P_\n-\d_{\n \r}P_\m), \;[L_{\m\n},K_\r]=i(\d_{\m \r} K_\n-\d_{\n \r}K_\m).  \label{algebra}
\end{align}

	The operators in a conformal field theory are divided into primaries and their descendants. Acting on a primary operator $O_{\D,l} (x)$  of scaling dimension $\D$ and spin $l$ ($\D=1$ for a free scalar field in four dimensions) we get the following rules

\begin{align}
&[D,O_{\D,l}]=i(x\cdot \p+\D)O_{\D,l}, \; [P_\m,O_{\D,l}]=i\p_\m O_{\D,l} \nonumber\\
&[L_{\m\n},O_{\D,l}]=i(x_\m \p_\n -x_\n \p_\m)O_{\D,l} +S_{\m\n} O_{\D,l} \nonumber\\
& [K_\m,O_{\D,l}]=i(x^2 \p_\m-2x_\m x\cdot \p-2\D x_\m)O_{\D,l}-2x^\n S_{\m\n} O_{\D,l},  
\end{align}
where $S_{\m\n}$ is the spin operator. Here we are using an index-free notation, where all the indices are contracted with a null vector $u^2=0$ (see section \ref{sec:primaries})

\begin{equation}
O_{\D,l}\equiv u^{\m_1}u^{\m_2} \dots u^{\m_l} O_{\D,\m_1 \m_2 \dots \m_l}.
\end{equation}

 Every primary operator corresponds to the lowest weight state of an irrep of the conformal algebra (\ref{algebra}). The descedants are then generated by acting with $P_\m$ on the primary operator. For each primary in the free theory there is an operator $V_{\D',l}$ in the interacting theory\footnote{The inverse is not true because the interacting CFT lives in non-integer dimensions and new operators appear \cite{Hogervorst:2015akt} that don't exist in integer dimensions.}  which in the $\a\rightarrow 0$ limit 

\[\lim_{\a\rightarrow 0} V_{\D',l} \rightarrow  O_{\D,l}, \]
and

\[\D'=\D+\g_O,\]
where $\g_O$ is the anomalous dimension of this operator. In general, there are many operators with the same scaling dimension and spin, and thus we have to write $O^i_{\D,l}$ with an extra index to take into account all these different operators. This degeneracy complicates things considerably, and in this section we focus on the non-degenerate case and we come back to this issue in section \ref{sec:generalization}. In this notation, the equation of motion $\Box \phi= \a \phi^3$ reads

\[ \Box V_{1,0}= \a V_{3,0}, \label{eq of motion}\]
where $\a$ is an unknown constant that can be determined from conformal symmetry alone \cite{Slava} to be $\frac{9}{2}\e$. Extending the discussion in that paper, we can use this constraint and the structure of the OPE between $V_{\D',l}$ and some other suitable operator to calculate the leading term of its anomalous dimension by doing only free-theory calculations.

In this work we consider only operators with twist $(\D-l)$ equal to the number of fields they contain and we can therefore exchange the index $\D$ with the number of fields $n$. The main idea is as follows: in the free theory both $O_{1,0}=\phi$ and $O_{3,0}=\phi^3$ are primaries and appear independently in the following OPE

\[O_{n-1,0}(x)O_{n,l}(0) \supset f\frac{(u\cdot x)^l}{|x|^{2(n-1+l)}}\( O_{1,0}(0)+\rho |x|^2 O_{3,0}(0)\) ,\label{first OPE}\]
where we have kept only the leading order in $x$. The constants $f$ and $\rho$ are determined by contracting the fields in these operators.  The three-point functions with the primaries $O_{1,0}$ and $O_{3,0}$ are therefore

\[ \<O_{n-1,0}(x)O_{n,l}(0)O_{1,0}(z)\>\approx f\frac{(u\cdot x)^l}{|x|^{2(n-1+l)}} \<O_{1,0}(0)O_{1,0}(z)\>, \nn\]

\[ \<O_{n-1,0}(x)O_{n,l}(0)O_{3,0}(z)\>\approx f \rho \frac{(u\cdot x)^l}{|x|^{2(n-2+l)}} \<O_{3,0}(0)O_{3,0}(z)\>, \label{free OPE}\]
where again we have kept only the leading order term in $x$. However, the corresponding operators $V_{1,0}$ and $V_{3,0}$ in the interacting theory are not independent anymore and they are related through the equation of motion (\ref{eq of motion}). In fact, the latter is a descendant of the former and enters in any OPE
only as $\Box V_{1,0}$. Hence, it appears in a way that is completely specified by conformal symmetry. The contribution of $V_{1,0}(0)$ in the interacting OPE is

\[V_{n-1,0}(x)V_{n,l}(0) \supset f'\frac{(u\cdot x)^l}{|x|^{\D'_{n-1,0}+\D'_{n,l}-\D'_{1,0}+l}}\(1+r \;x\cdot \p +q\; x^\m x^\n\p_\m\p_\n+p\;x^2 \Box+\dots \)V_{1,0}(0) \label{int OPE},\]
where the constants $r,q$ and $p$ are fully determined in terms of the $\D_n$'s and $l$ (see Appendix \ref{sec:OPE}). The only important constant here is $p$ because this term will give rise to the $\phi^3$ term in the free theory and is given by

\[p=\frac{(\D_{n-1,0}+\D_{1,0}-\D_{n,l}+l)(\D_{n-1,0}-\D_{1,0}-\D_{n,l}-l)}{16\D_{1,0}(\D_{1,0}+1)(\D_{1,0}-\d)} \label{p coeff},\]
where $\d=1-\frac{\e}{2}$ is the dimension of the free field $\O_{1,0}=\phi$. The counterparts of the correlators (\ref{free OPE}) in the interacting case are

\[ \<V_{n-1,0}(x)V_{n,l}(0)V_{1,0}(z)\>\approx f' \frac{(u\cdot x)^l}{|x|^{\D'_{n-1,0}+\D'_{n,l}-\D'_{1,0}+l}} \<V_{1,0}(0)V_{1,0}(z)\>, \label{V1}\]
and

\[ \<V_{n-1,0}(x)V_{n,l}(0)V_{3,0}(z)\>\approx f' \frac{(u\cdot x)^l}{|x|^{\D'_{n-1,0}+\D'_{n,l}-\D'_{1,0}+l}} \<\a\;p\; V_{3,0}(0)V_{3,0}(z)\>, \label{2}\]
where we have used the OPE the constraint (\ref{eq of motion}). Requiring that these correlators should approach those of the free theory in the $\a\rightarrow 0$ limit, we are led to the conditions

\[f'\rightarrow f, \quad \a p\rightarrow \rho. \label{conditions}\]

The crucial observation here is that the denominator of (\ref{p coeff}) in the $\e\rightarrow 0$ limit is proportional to the anomalous dimension of $V_{1,0}$ which was shown in \cite{Slava} to be quadratic in $\e$.       More precisely it was shown that

\[\g_{1,0}=  \frac{1}{108}\e^2. \]

 Since $\rho$ is finite we get a non-trivial condition for the numerator of (\ref{p coeff}) and therefore an equation  for $\D'_{n,l}$ in terms of $\g_{n-1,0}$, which was also computed in the same paper to be 

\[\g_{n,0}=\frac{1}{6}n(n-1)\e. \label{g(n,0)}\]
More specifically, substituting $\D'_{n,l}=n\d+l+\g_{n,l}$ in (\ref{p coeff}) and keeping only the leading order terms in $\e$ we get

\[p=\frac{l+1}{16}\frac{\g_{n,l}-\g_{n-1,0}-\g_{1,0}}{\g_{1,0}},\]
and using (\ref{conditions}) we conclude that for $n>2$

\[\g_{n,l}=\(\frac{1}{6}(n-1)(n-2) +\frac{2}{3}\frac{\rho}{l+1}\) \e, \label{main}\]
where we also used $\a=\frac{2}{9}\e$. For $n=2$ we have to contract two fields from $O_{3,l}$ with $O_{2,l}$ in (\ref{free OPE}), which is zero. In this case $\r=0$ and therefore $p$ should be just finite implying only that $\g_{2,l}\sim \e^2$, without giving any non-trivial condition for the coefficient in front. However, a modification is possible in order to include the case of two fields (see section \ref{sec:currents}).

In conclusion, we can use (\ref{main}) to calculate the anomalous dimension of non-degenerate operators with $n>2$ by only evaluating $\rho$ in the free theory. An example is the primary operators with spin two. These are non-degenerate and we can calculate their anomalous dimensions in this way. They are given by
\[O_{n,2}=\phi^{n-2} T=\phi^{n-2} (\frac{1}{2}\phi \p^2 \phi-\p\phi\p\phi) ,\] 
where $T$ is the energy momentum tensor (\ref{energy momentum tensor}) and $\p \equiv u \cdot \p$. In this case $\r=n-2$ and therefore

\[\g_{n,2}=\frac{1}{6}(n-2)(n-5)\e. \]
Another example are the operators with spin three which are given by
\[O_{n,2}=\phi^{n-2} \(\p\phi T -\frac{1}{6} \phi \p T\).\] 
In this case $\rho=\frac{1}{5}(n-4)$ and therefore
\[\g_{n,3}=\frac{1}{6}\((n-2)n-2\)\e. \]

We see that to apply this method, we first need to find a way to calculate the coefficients $f$ and $\r$ in the free theory, and in addition we also need to deal with the degenerate cases. In the following section, we describe a way to construct the class of primary operators considered here, and in the subsequent sections we will deal with the latter issue.

\section{Primaries of the free massless scalar theory in d=4.\label{sec:primaries}}

In this section, we construct explicitly a class of operators, namely those whose twist is equal to the number of fields they contain. The problem of determining the operator spectrum of a CFT has been addressed many times in the literature. For the case of two fields and three fields there are known expressions \cite{Makeenko:1980bh,Kehrein:1992fn,Braun:1998id,Braun:1999te,Braun:2001qx,Kehrein,Mikhailov:2002bp}, see also \cite{Braun:2003rp} for a nice review. Here our aim is to reconstruct those expressions with a slightly different method which can be easily extended to the operators with more fields.

 In the four-dimensional free scalar CFT, a primary operator is a symmetric and traceless combination of terms like

\[ \p_{\m_1 \m_2 \m_3 \dots} \p_{a} \phi \; \p_{\n_1 \n_2 \n_3 \dots} \p^{a}\phi \;\p_{\k_1 \k_2 \k_3 \dots} \phi \; \dots
\label{primaries}\]
where $\Box\phi=0$. Generally, a symmetric traceless tensor $f_{\m_1 \m_2 \dots \m_l} (x)$ can be encoded in a generating functional \cite{SPIN,Costa:2011dw}

\[f(u,x)=u^{\m_1} \dots u^{\m_l} f_{\m_1 \dots \m_l} (x).\]
where $u$ is a null vector ($u^2=0$). Using the differential operator \cite{Dobrev:1975ru}

\[D_\m=(\frac{d}{2}-1+u\cdot\frac{\p }{\p u})\frac{\p }{\p u^\m}-\frac{1}{2}u_\m\frac{\p^2 }{\p u\cdot \p u}. \label{D operator}\]
the tensor can be recovered from
\[ f_{\m_1 \m_2 \dots \m_l} (x)= \frac{1}{s!} D_{\m_1}D_{\m_2} \dots D_{\m_l}f(u,x) \label{sym tensor},\] 
The operator (\ref{D operator}) has the properties

\[ [D_\m,D_\n]=0, \quad D^\m D_\m=0.\]
The idea behind this trick is that this operator has an additional property when acting on a function proportional to $u^2$

\[ D_\m u^2 f(u,x)=u^2 g_\m(u,x),\]
where $f(u,x)$ is an arbitrary function of $u$ and $x$. The right hand side depends on the derivatives of $f(u,x)$, but its exact form is irrelevant here. However, the crucial point is that the right-hand side is still proportional to $u^2$ and therefore, we can set it equal to zero even before we apply the $D_\m$'s in
(\ref{sym tensor}).

The primary operators of the free scalar theory (\ref{primaries}) are built from the objects

\[\p^N \phi(x) \equiv (u\cdot\frac{\p}{\p x})^N \phi(x)=(-i u\cdot P)^N\phi(x).\]
 The generating functional for the primaries with $M$ fields has the form

\[ B_{M,l}(x) =\sum_{ N_1+N_2+\dots=l} g(N_1,N_2 \dots,N_M)\; \p^{N_1} \phi \p^{N_2} \phi \dots \p^{N_M} \phi(x) \label{gen fun},\]
where $g(N_1,N_2 \dots,N_M)$ is an appropriate function so that this combination satisfies

\[ K_\m B_{M,l}(0)=0. \label{def of prim}\]
Here we denote these operators by $B_{M,l}$. As we will see shortly in most cases these operators are degenerate and therefore, these expressions are just a particular basis.

Having constructed this generating functional, we can apply the $D_\m$ operator $l$ times and get the primary operators with $l$ indices. Hence, the problem of determining the primary operators has been reduced to the problem of finding the appropriate function $g(N_1,N_2 \dots,N_m)$ that ensures (\ref{def of prim}). The way we proceed is first to determine the action of $K_\m$ on $\p^{N} \phi$, and then apply it on (\ref{gen fun}) to find a recursion relation for the coefficient $g(N_1,N_2 \dots,N_l)$. Using $K_\m\phi(0)=0$, we have that

\[ K_\m \p^{N} \phi= [K_\m,(-i u\cdot P)^N
] \phi.\]
This commutator is easy to calculate using the conformal algebra (\ref{algebra}):

\begin{align}
[K_\m,(-i u\cdot P)^N]&= -(i u\cdot P)^{N-1} \{2N(u^\n L_{\m\n}+u_\m D)+2iN(N-1)u_\m\}.
\end{align}
Therefore, for a scalar field $\phi$ of scaling dimension $\D$ satisfying $L_{\m\n} \phi=0$

\[ K_\m \p^{N} \phi(0)=-2iu_\m \(\D N+N(N-1)\)\p^{N-1} \phi(0). \]
For a free scalar field of dimension $\D=1$ (free scalar theory in $d=4$) we have

\[ K_\m \p^{N} \phi(0)=-2iu_\m N^2\p^{N-1} \phi(0). \label{K}\]
We can now use this rule and determine the coefficient in (\ref{gen fun}) to construct the generating functions which from now on we refer to as primaries, since we are not going to use operators with restored indices. The primary with one field is just
\[B_{1,0}=\phi.\] The primary with two fields will be of the form
\[B_{2,l}=\sum_{N=0}^l f_N \p^{l-N} \phi \p^{N} \phi.\]
Acting with $K_\m$ on two adjacent terms we get

\begin{align}
& f_N \{ (l-N)^2 \p^{l-N-1} \phi \p^{N} \phi+N^2\p^{l-N} \phi \p^{N-1} \phi \} + \nn \\
& f_{N+1} \{ (l-N-1)^2 \p^{l-N-2} \phi \p^{N+1} \phi+(N+1)^2 \p^{l-N-1} \phi \p^{N} \phi \}.
\end{align}
In order to cancel all these terms, $f_N$ has to satisfy the following recursion relation

\[ (l-N)^2 f_N+(N+1)^2 f_{N+1}=0.\]
This can be done only for even $l$ (otherwise we cannot cancel the last term $\p^{(l-1)/2}\phi \p^{(l-1)/2}\phi$) and the solution is given by

\[f_N=\frac{(-1)^N}{(l-N)!^2 N!^2}.\]
Hence, we conclude that the primaries with two fields necessarily have an even number of derivatives and are given by

\[B_{2,l}=\sum_{N=0}^l \frac{(-1)^N}{(l-N)!^2 N!^2} \p^{l-N}\phi \p^N \phi. \label{currents}\]

As an example, the primary with two fields and two derivatives is 

\[T=\frac{1}{2} \phi \p^{2} \phi -\p \phi\p \phi, \label{energy momentum tensor}\]
and acting twice with the $D$ operator (\ref{D operator}) we get
\[ -\frac{1}{4} D_{\m}D_{\n} B_{2,2}=\p_{\m}\phi \p_{\n}\phi +\frac{1}{2}\phi\p_{\m}\p_{\n}\phi- \frac{1}{4}\d_{\m\n}\p\phi \cdot \p\phi, \]
which is the improved energy-momentum tensor.

Moving to the primary operators with three fields, we can start from the general form 

\[B_{3,l}=\sum_{N_1,N_2} f_{N_1,N_2} \p^{l-N_1-N_2}\phi \p^{N_1}\phi \p^{N_2}\phi,\]
but it is much more convenient to write the last two factors in terms of the primaries with two fields and their derivatives. Denoting the N'th derivative of the primary $B_{M,l}$ as

\[\p^N B_{M,l},\]
we have that
\[K_\m \p^N B_{M,l}= [K_\m,(-iu\cdot P)^N]B_{M,l}=-2iN\(2l+M+N-1\)u_\m \p^{N-1} B_{M,l}.\]
Now we can construct the primary operators with three fields and $l$ derivatives as

\begin{align}
& \phi B_{2,l} \nn \\
& \phi \p^2 B_{2,l-2}+ a \p\phi \p B_{2,l-2} +b \p^2\phi B_{2,l-2} \nn \\
&\dots,
 \end{align}
where $a$ and $b$ are constants (like $f_N$). Here we have assumed that $l$ is even (in the opposite case we start with $\phi\p B_{2,l-1}+a \p\phi B_{2,l-1}$). Thus, we see that we can construct many operators with the same spin as

\[B_{3,l}^P=\sum_{N=0}^P f_N^P \; \p^{P-N} \phi \p^N B_{2,l-P} ,\]
where again $f_N^P$ is a constant to be determined by the requirement that these operators are annihilated by $K_\m$. The extra index $P$ labels all the different operators we can construct for the lase $l$. In this case the recursion relation that we get is

\[ (P-N)^2 f_N+(N+1)(2l-2P+N+2) f_{N+1}=0,\]
and has following solution

\[f_N^P=\frac{(-1)^N}{(P-N)!^2 N! (2l-2P+N+1)!}.\]
So, the primary operators with three fields are given by
\[B_{3,l}^P=\sum_{N=0}^P \frac{(-1)^N}{(P-N)!^2 N! (2l-2P+N+1)!} \; \p^{P-N} \phi \p^N B_{2,l-P} \label{three fields},\]
where for even $l$
\[P=0,2,\dots,l/3, \]
while for odd $l$
\[P=1,3,\dots,l+3/3. \]

The possible values for $P$ were determined by noting that (see Appendix \ref{sec:Operator counting}) the number of primaries increases by one when $l$ increases by 6. One may wonder if the independent operators are the ones that appear first in the sequence. The answer to this question is that every time we increase $P$, new terms appear when it is less than $l/3$ (think of the distribution of derivatives among the $\phi$'s) and therefore the first values of $P$ give independent operators. Then, we can proceed recursively and construct the primary operators with M fields as follows

\[B_{M,l}^P=\sum_{N=0}^P \frac{(-1)^N}{(P-N)!^2 N! (2l-2P+N+M-2)!} \; \p^{P-N} \phi \p^N B_{M-1,l-P}   \label{primary op},\]
 where we have suppressed the index in $B_{M-1,l-P}$ that takes into account the different operators at this level. These operators are not independent as we show now. The number of primary operators $C_{M,l}$, whose twist is equal to the number of fields, is shown in Appendix \ref{sec:Operator counting} to be given by

\[\sum_l  C_{M,l} \; q^l=\prod_{n=2}^M \frac{1}{(1-q^n)}.\]
However, when $M>3$, for each $B_{M-1,l}$ operator we can construct $l$ operators $B_{M,l}$ for all the different values of $P$. These operators are not all independent because after expanding the above expression for the $M=3$ and $M=4$ cases we get

\[M=3: \quad 1 + q^2 + q^3 + q^4 + q^5 + 2 q^6 + q^7 + 2 q^8 + 2 q^9+ \cdots \nn \]

\[M=4: \quad 1 + q^2 + q^3 + 2 q^4 + q^5 + 3 q^6 + 2 q^7 + 4 q^8+ 3 q^9 +\cdots \nn \]
and we see that for $B_{4,2}$ we get only one operator, in contrast to what one would expect using $B_{3,0}$ and $B_{3,2}$. Probably there is an upper limit for the index $P$ as in the $M=3$ case, but we couldn't find a nice systematic way to determine it. Note also that these operators are neither normalized nor orthogonal.

With a little bit more work we can extend these expressions to arbitrary $d \geq 4$ dimensions. The results are

\[ B_{2,l}=\sum_N \frac{(-1)^N}{N! (\frac{d-4}{2}+N)!(l-N)!(\frac{d-4}{2}+l-N)!} \p^N \phi \p^{l-N} \phi, \]
and 

\[B_{M,l}^P=\sum_{N=0}^P \frac{(-1)^N}{(P-N)!(\frac{d-4}{2}+P-N)! N! (2l-2P+N+\frac{d-2}{2}M-\frac{d}{2})!} \; \p^{P-N} \phi \p^N B_{M-1,l-P} \nn .\]

We conclude this section be noting that these operators are not the only primary operators of the theory. There exist primary operators whose twist is greater that the number of fields. The first one appears for $\D= 2d $ which is the energy momentum squared $T_{\m\n} T^{\m\n}$. In general, the full operator spectrum will consist of appropriate combinations of terms like

\[ \Box^{n_1} \p^{m_1} B_{M_1,l_1}\Box^{n_2} \p^{m_2} B_{M_2,l_2} \dots , \label{other primaries}\]
and one can try to repeat the above construction, but this is beyond the scope of this work.

\section{Anomalous dimensions of weakly broken currents \label{sec:currents}}

In this section, we modify the method described in the previous sections, and we rederive the well-known leading order term in the anomalous dimensions of operators with two fields. In the free theory these operators are conserved currents,\footnote{These operators saturate the unitary bounds imposed by the conformal algebra \cite{Mack:1975je,Minwalla} and therefore they have to obey the conservation equation} and are important for applications to AdS/CFT  correspondence \cite{Klebanov:2002ja}.

As mentioned before, since the OPE coefficient $\rho$ is zero, we cannot apply the method described in section \ref{sec:method} to calculate their anomalous dimensions. However, we can extend this method using the fact that in the free theory these operators obey the conservation equation, which in the interacting theory will get corrections on the right-hand side \cite{Belitsky}

\[ \p\cdot D V_{2,l} (0)= \a K_{l-1} \label{non conserv},\]
where $D$ is the operator defined in (\ref{D operator}) and $K_{l-1}$ is a descendant  operator of spin $l-1$. 
We start with equation (\ref{concrete OPE})

\[V_{1,0}(x) V_{2,l}(0) \supset f'\frac{(u\cdot x)^l}{|x|^{\D'_l+l}} C(x,\p)V_{1,0}(0). \]
Applying the operator $\p\cdot D$ we can write it as

\[\a V_{1,0}(x) K_{l-1} \supset f'\frac{(u\cdot x)^{l-1}}{|x|^{\D'_l+l}} C'(x,\p)V_{1,0}(0) ,\]
where

\[C'(x,\p)=f'(t'_0+p'_0 x^2 \Box+\dots),\]
with known coefficients $t'_0$ and $p'_0$, whose $\a \rightarrow 0$ limit is (see appendix \ref{sec:C'})

\[t'_0 \rightarrow 0 ,\nn \]
\[p'_0 \rightarrow  -\frac{27}{4} l (1 + l)^2 \frac{\g_l}{\e^2}\nn .\]
Following the same steps as before, we arrive at

\[(u\cdot x)\frac{\<V_{1,0}(x)K_{l-1}(0)V_{3,0}(z)\>}{\<V_{1,0}(x)V_{2,l}(0)V_{1,0}(z)\>}= \frac{3}{4}(t'_0+8p'_0 \frac{x^2}{z^2}+\dots)\frac{1}{z^2} \label{curr exp}.\]
Thus, we see that in the $\a \rightarrow 0$ limit only the second term survives, giving a non-trivial condition for the anomalous dimension of these operators. In this limit, the above three-point functions can be calculated to leading order, using the operators of the free theory. The leading term in the denominator is coming from the term $\phi \p^l \phi$ in $B_{2,l}$, while the leading term in the numerator is coming from the term $\phi^3 \p^{l-1} \phi$ in $K_{l-1}$. From (\ref{currents}) we see that

\[B_{2,l}=\frac{2}{l!^2} \phi \p^l \phi +\dots ,\]
while using (\ref{non conserv}) one can get 

\[K_{l-1}= -\frac{1}{(l-1)!^2}\frac{(l+3)(l+1)(l-2)}{2l}\phi^3 \p^{l-1} \phi +\dots .\]
Using (\ref{curr exp}) we obtain

\[-\frac{1}{8}(l+3)(l+1)(l-2)=p'_0,\]
and plugging the expression for $p'_0$ we arrive at

\[\g_l= \frac{1}{54} \(1-\frac{6}{l(l+1)}\)\e^2 .\]
This result is well known for many years and was rederived in \citep{Anselmi:1998ms,Belitsky,Giombi} in a similar way.

\section{Generalization \label{sec:generalization}}

In the previous section we calculated the anomalous dimensions of operators $V_{2,l}$, which are non-degenerate primaries. However, the operators made of three fields are degenerate for spin larger that five. This is a known fact \cite{Kehrein} and in our notation is reflected by the additional index $P$ in (\ref{three fields}). Also, the operators in (\ref{three fields}) are not orthogonal\footnote{Nevertheless, there are known orthonormal expressions for operators with three fields \cite{Braun:2003rp}.}, but even if they were, we could always rotate them and get a new set of primary operators, and we don't know a priori which orthogonal set to use.

To illuminate this more, consider an orthonormal set of free primary operators $O^i_{n,l}$ with the same scaling dimensions and spin. Conformal symmetry dictates 

\[ \<O^i_{n,l}(x) O^j_{n,l}(x)\>= \delta^{ij} \frac{(u I u )^l}{|x|^{2(n+l)}}, \]
where

\[ I_{\m\n}= \delta_{\m\n} -2 \frac{x_\m x_\n}{x^2}. \] 
Here we have set equal the auxiliary vectors for each operator. In general, the primary operators of the interacting theory, $V^i_{n,l}$, will approach a rotated set of the above operators
\[ \lim_{\a\rightarrow 0} V^i_{n,l}= \sum_j R^i\!_j O^j_{n,l} ,\]
where $R$ is an unknown rotational matrix. Hence, even if we had such a set of operators, we don't know their $\a\rightarrow 0$ limit, because we cannot determine the matrix R from the free theory.

 Fortunately, we can generalize the method described in the previous sections using any set of primary operators (not necessary orthonormal). We begin by writing (\ref{int OPE}) as 

\[V_{n-1,0}(x)V_{n,l}(0) \supset f'\frac{(u\cdot x)^l}{|x|^{\D'_{n-1,0}+\D'_{n,l}-\D'_{1,0}+l}} C(x,\p)V_{1,0}(0) \label{concrete OPE},\]
where the derivatives on the right-hand side are with respect to the argument of $V_{1,0}$, which is afterwards set to zero. From equations \ref{V1} and \ref{2} we get

\[\<V_{n-1,0}(x)V^i_{n,l}(0) V_{3,0}(z)\> \sim 9(l+1) p_i \frac{x^2}{z^4} \<V_{n-1,0}(x)V^i_{n,l}(0) V_{1,0}(z)\>  \label{generalization}.\] 
where $\sim$ means that the leading order terms on both sides are equal and 
\[p_i=  \frac{\g^i_{n,l}-\g_{n-1,0}}{\e}. \label{generalized main}\]
 
 For every complete set of primary operators (for instance $B_{M,l}$)
\[ \lim_{\a\rightarrow 0} V^i_{n,l} = \sum_\b L_{ij} O^j_{n,l} ,\]
where $L$ is an invertible matrix (not necessary orthogonal).

Defining a new matrix $P_{\a\b}$ in the free theory as 
\[\<\phi^{n-1}(x)B^i_{n,l}(0) \phi^3(z)\> \sim  P^i\!_{j} \frac{x^2}{z^4} \<\phi^{n-1}(x)B^j_{n,l}(0) \phi(z)\>  ,\label{P matrix}\]
and multiplying both sides by $L$ 

\[\<\phi^{n-1}(x)L_{ij}B^j_{n,l}(0) \phi^3(z)\> \sim L_{ij}P^{jk}L^{-1} _{k m} \frac{x^2}{z^4} \<\phi^{n-1}(x)L^{m}\!_q B^q_{n,l}(0) \phi(z)\> , \nn\]
we see that its eigenvalues are equal to

\[9(l+1)p_i. \]

 However, equation (\ref{P matrix}) can determine at most one eigenvalue and that's only if the two vectors on both sides are the same. On the other hand, we could replace $\phi^{n-1}$ with $V_{n-1,l'}$, and find other equations of this form and by using those we could determine all the eigenvalues of $P$. Here we will show that this is possible for $n=3$, and in the next section we will employ this technique and calculate the anomalous dimensions of operators with three fields.
Consider the analogous equation of (\ref{generalization}), where now the primary with two fields has non-zero spin

\[\<V_{2,l'}(x)V^i_{3,l}(0) V_{3,0}(z)\> \sim h^i \frac{x^2}{z^4} \<V_{2,l'}(x)V^i_{3,l}(0) V_{1,0}(z)\> . \label{h first}\]
for some constant $h^i$. As in the $l'=0$ case, we can define the matrix $P$ as

\[ \vec{Y}^{(3)}_{l' l}  \sim P \cdot \vec{Y}^{(1)}_{l' l} \label{Y},\]
where 

\begin{align*}
&\vec{Y}^{(3)}_{l' l}=\<B_{2,l'}(x)\vec{B}_{3,l}(0) \phi^3(z)\> , \\
&\vec{Y}^{(1)}_{l' l}=\frac{x^2}{z^4}\<B_{2,l'}(x)\vec{B}_{3,l}(0) \phi(z)\>  .
\end{align*}
The operators $B_{M,l}$ are those we constructed in section \ref{sec:primaries}, and therefore this matrix can be calculated from the free theory and its eigenvalues are equal to $h^i$.

Moreover we argue that $h^i$ must be of the form

\[h^i = c_{l'}\frac{\g^i_{3,l}-\kappa_{l'}\g_{2,l'}}{\e}. \label{h}\]
Looking at (\ref{GF exp}) we see that

\[ (\D_1-\d) p=\Gamma       ,\]
where $\Gamma$ is a non-singular (in $\e$) rational function of the $\D$'s and $l$'s. Demanding that (\ref{h first}) has the correct free theory limit

\[h^i \propto  \a p \propto \frac{\Gamma}{\e},\]
and since $h^i$ has to be regular, $\Gamma$ must be proportional to the anomalous dimensions of $V_{3,l}$. In other words

\[\Gamma \propto \g^i_{n,l} -\kappa_{l'}\g_{n-1,l'},\]
and therefore we arrive at (\ref{h}). The constants $c_{l'}$ an $\kappa_{l'}$ cannot be fixed for a generic value of $l'$ and reflect the multiple conformal structures we can built (see Appendix \ref{sec:T ope}). Nevertheless, for $l'$ equal to zero or two we have

\begin{align}
& l'=0: \quad c_{l'}=9(l+1), \; \kappa=1 \\
& l'=2: \quad c_{l'}=\frac{ (l+1)^2 }{3(l+3)}, \; \kappa=0 
\end{align}
Note that for $l'=2$ the operator $V_{2,2}$ is the energy momentum tensor and therefore has zero anomalous dimensions.

\section{Operators with three fields\label{sec:three fields}}

In this section, we calculate the leading order term in the anomalous dimensions of operators with three fields. Apart from the traditional diagramatic way, these have also been calculated using conformal symmetry \cite{Kehrein,Derkachov:1995wg} but in a different way. The purpose of this section is therefore to rederive those results using the method in \cite{Slava}. In the free theory these operators are given by (\ref{three fields}), and for spin greater than five are degenerate. Therefore, in the interacting theory these operators get mixed and we have to follow the method described in the previous section.

Our starting point is equation (\ref{Y}). We choose the operators $O^i$ to be the ones we constructed in section \ref{sec:primaries}. Looking at the expression for $B_{3,l}$ for even spin, in (\ref{three fields}), we see that they are of the form

\[B^i_{3,l}= \frac{2}{(2l-i+1)! (l-i)!^2 i!}\phi^2 \p^l \phi + \dots.\]

Since only this term will contribute at leading order in $\<\phi^2(x)B^i_{3,l} \phi^3(z)\>$, we immediately deduce that

\[ \<\phi^2(x)B^i_{3,l} \phi^3(z)\>= 24\frac{ 2^l l!}{(2l-i+1)! (l-i)!^2 i!} \frac{(u \cdot x)^l}{x^{2(l+1)}} \frac{1}{z^6}.\]

The other three-point function needs just a little bit more work. First we observe that for the leading order term we only need the terms in $B_{3,l}$ with at least one $\phi(0)$ without derivatives, to be contracted with $\phi(z)$. Therefore the important terms are

\[B^i_{3,l}=\frac{2}{(l-i)!^2} \sum_{N=0} ^i f_N^i\; \phi \;\p^{i-N} \phi \; \p^{l-i+N} \phi + \dots.\]
Doing the contractions and performing the sum we finally get

\[ \<\phi^2(x)B^i_{3,l} \phi(z)\>=   \frac{ 2^{2l +2} l!}{(2l-i+1)! (l-i)!^2 i!} \frac{(u \cdot x)^l}{x^{2(l+2)}} \frac{1}{z^2}.\]
So, we see that by rescaling the operators

\[ B^i_{3,l} \rightarrow  \frac{(2l-i+1)! (l-i)!^2 i!}{ 2^l l!}B^i_{3,l}  ,\]
we get

\[ \<\phi^2(x)B^i_{3,l} \phi^3(z)\>= 24\frac{(u \cdot x)^l}{x^{2(l+1)}} \frac{1}{z^6} \nn,\]
\[ \<\phi^2(x)B^i_{3,l} \phi(z)\>=4 \frac{(u \cdot x)^l}{x^{2(l+2)}} \frac{1}{z^2},\] 
independently of $i$. Hence, the eigenvalue equation (\ref{Y}) becomes 
 
 \[ P \cdot \(\begin{matrix}
           1 \\
           1 \\      
           \vdots \\
           1
         \end{matrix} \)= \frac{2}{3(l+1)}\(\begin{matrix}
           1 \\
           1 \\           
           \vdots \\
           1
         \end{matrix} \) ,\]
which implies that the $P$ matrix has an eigenvalue equal to $2/3(l+1)$, and therefore we conclude that there is an operator with anomalous dimension

\[\g_{3,l}=\frac{1}{3}\(1+\frac{2}{l+1}\)\e.  \]

For odd $l$ one can repeat the same steps and arrive at a similar result that differs only by a relative minus sign. Combining both results we can write

\[\g_{3,l}=\frac{1}{3}\(1+(-1)^l\frac{2}{l+1}\)\e. \label{g3} \]

Moving forward we repeat the same procedure for $l' \neq 0$. 
Instead of using $P$, for this case, it is more convenient to promote $\g_{3,l}$ into a matrix whose eigenvalues are the anomalous dimensions of operators with three fields. We also go to another basis formed by $B'^0=B^0$
and 

\[ B'^i =B^i -B^0, \quad i \neq 0,\]
where the operators appearing on the right-hand side are the rescaled operators we used previously. In this new basis, we already know that 

 \[ \Gamma_{3,l} \cdot \(\begin{matrix}
           1 \\
           0 \\      
           \vdots \\
           0
         \end{matrix} \)= \frac{1}{3}\(1+(-1)^l\frac{2}{l+1}\)\e \(\begin{matrix}
           1 \\
           0 \\           
           \vdots \\
           0
         \end{matrix} \) \label{1}.\]
 Next, we notice that since the term $\phi^2 \p^ l\phi$ is absent from the $B'$'s, for all the values of $l'$ we have that

 \[\vec{Y}^{(3)}_{l' l}= \(\begin{matrix}
           a_{l'} \\
           0 \\
           0\\
           \vdots \\
           0
         \end{matrix} \), \label{Y3}\]
 for some constant $a_{l'}$. Here we have gotten rid of the $x$ and $z$-dependence since it is the same on both sides of (\ref{Y}). On the other hand, $\vec{Y}^{(1)}_{l' l}$ will be different in general. For instance, one can easily evaluate $\< B_{2,2}(x) B^i_{3,l}(0) \phi(z) \>$ and check that you get a different vector, whose actual expression is not essential here. Therefore, we conclude that in addition to (\ref{1}), the set of equations for all the values of $l'\neq 0$ will take the form

\[ \Gamma_{3,l} \cdot \(\begin{matrix}
           1 \\
           0 \\
           0\\
           \vdots \\
           0
         \end{matrix} \)=\g_{3,l} \vec{X}_{l'},\]
         
where $\vec{X}_{l'}$ is the vector we get for each value of $l'$ rescaled by $a_{l'}$. Since for arbitrary $l'$ the vector $\vec{X}_{l'}$ is not parallel to (\ref{Y3})\footnote{It has been checked numerically for $l<400$ that the first vectors for $l'<l/3$ are independent.}, after subtracting (\ref{1}) from the above equation, we conclude that $\Gamma_{3,l}$ has only one non-zero eigenvalue. This implies that among the operators with three fields, only one acquires anomalous dimension at order $\e$ which is given by (\ref{g3}). This matches exactly the results in \cite{Kehrein:1992fn,Kehrein}, where the same results were obtained, but in different way. Note that this implies that at one loop order the primary operators are
$ B'_0+ b_i B'^i$ and $ S^i \! _j B'^j $ with $i,j \neq 0$, and where $\vec{b}$ and $S$ are an unknown vector and matrix respectively, which give us an additional freedom in choosing a different operator basis.

One could try to repeat the same for the operators with four fields, but it turns out that the matrix of anomalous dimensions is not fixed completely because of the unknown constants $\kappa_{l'}$, which reflect the different conformal structures. Nevertheless, we can still express the anomalous dimensions and the constants $c_{l'}$ in terms of the $\kappa$'s.

\section*{Acknowledgments}

I would like to thank Leonardo Rastelli for suggesting the project and for helpful discussions during all the stages of this work. I'm also very grateful to Martin Ro\v{c}ek for fruitful conversations and for various comments on the manuscript. This work was supported by NSF Grant No. PHY-1620628.

\section*{APPENDICES}
\appendix

\section{Operator Counting \label{sec:Operator counting}}

In this appendix we review how to count the primary operators that are constructed in section \ref{sec:primaries}.  We also give a more general way to determine the number of primary operators with scaling dimension $\D$ and spin $l$ based on group theory\footnote{We are grateful to Leonardo Rastelli for pointing out this method.}. 

Let's consider the terms in the operators (\ref{primary op}). These are all the possible ways of acting with $l$ derivatives on $M$ fields

\[\phi^{M-1} \p^l \phi, \; \phi^{M-2} \p \phi \p^{l-1} \phi, \; \dots  .\]
The number of these terms is equal to the $M$-restricted partition of $l$ objects $p_M (l)$. It is a well-known fact from combinatorics that there is a generating functional for these number, given by

\[\sum_{l=0} ^{\infty} p_M (l)\; q^l = \prod_{n=1}^M \frac{1}{1-q^n} \label{partitions}.\]
However, these terms can be organized into linear combinations of the primary operators $B^P_{M,l}$ and their descendants (this is a change of basis), where $P$ labels the different operators. If the number of these operators is $C_{M,l}$, or in other words, if $i$ runs from one to $C_{M,l}$, we have the following relation

\[ C_{M,l}+C_{M,l-1}+ \dots +C_{M,0} =p_M(l), \] 
or equivalently 

\[ C_{M,l}=p_M(l)-p_M(l-1).\]
Multiplying by $q^l$ and summing over $l$ we get

\[\sum_{l=0} ^\infty C_{M,l} \; q^l=(1-q)\sum_{l=0} ^\infty p_M (l)=(1-q)\prod_{n=1}^M \frac{1}{1-q^n}.\]
Therefore, we arrive at the following generating functional for the number of primary operators, whose twist is equal to the number of fields,

\[\sum_{l=0} ^\infty C_{M,l} \;q^l=\prod_{n=2}^M \frac{1}{1-q^n}.\]
For the case of just three fields the generating functional becomes

\[\frac{1}{(1-q^2)(1-q^3)},\]
and expanding in Taylor series we see that the number of primary operators is increased by one when $l$ is increased by six.

In the rest of this appendix, we describe a more rigorous way, based on group theoretical arguments, for counting the number of all the primary operators of the theory. The Cartan subalgebra of the conformal algebra consists of the generators $D,J_1,J_2$ where the last two generators refer to the two $SU(2)$ subalgebras of the Lorentz group $SO(3,1)\approx SU(2)\times SU(2)$. Hence, each irrep of the conformal algebra is labeled with three numbers $(\D,j_1,j_2)$ corresponding to these three generators. Moreover, the conformal algebra has long and sort irreps. A long irrep contains all the states

\[P_{\m_1}P_{\m_2}\dots\ket{\D,m_1,m_2} \quad m_i=-j_i, \dots, +j_i,\]
 while a short irrep is constrained through a shortening condition. The free scalar theory has two such irreps, namely $(1,0,0)$ and $(2+l,\frac{l}{2},\frac{l}{2})$, or 

\[P^2 \phi(0) \ket 0=0  ,\]
and
\[P_\m J^\m(0) \ket 0=0 , \]
respectively. Using this labeling we can define the character of each irrep as

\[\x_{\D,j_1,j_2}(s,x,y)= \sum_{\D,m_1,m_2} s^{2\D} x^m_1 y^{m_2}.\]
For an irrep $(\D,j_1,j_2)$ this leads to \cite{Dolan:Characters}

\begin{align}
\x_{\D,j_1,j_2}(s,x,y)=&\(\sum_{m_1=-j_1}^{j_1} x^{m_1}\)\(\sum_{m_2=-j_2}^{j_2} y^{m_2}\) s^{2\D}\(1+\(xy+\frac{x}{y}+\frac{y}{x}+\frac{1}{xy}\)s^2 +O(s^4)\) \nn\\
=&s^{2\D} \x_{j_1}(x)\x_{j_2}(y)P(s,x,y), \nn
\end{align}
where

\[P(s,x,y)=\frac{1}{(1-xy s)(1-\frac{x}{y}s)(1-\frac{y}{x}s)(1-\frac{1}{xy}s)},\]
and 

\[\x_{j}(x)=\sum_{m=-j}^{j} x^{m}=\frac{x^{j+1/2}-x^{-j-1/2}}{x^{1/2}-x^{-1/2}}.\]
The latter are the characters of $SU(2)$ that obey the following orthogonality condition

\[\oint \frac{dx}{2\pi i} \frac{1-x}{x}\x_{j}(x)\x_{j'}(x)=\d_{jj'} ,\]
with the complex integral taken along the unit circle. Using the notation $\p_{ab}=\sigma^\m_{ab} \p_\m$, where the two Latin indices correspond to the two SU(2)'s, the operators in the irrep created by $\phi$ are

\[\phi, \; \p_{\pm\pm}\phi, \; \p_{\pm\pm}\p_{\pm\pm}\phi \dots .\]
After removing the null states

\[\Box\phi, \; \p_{\pm\pm}\Box\phi \dots,\]
the character or this irrep becomes 

\begin{align}
\x_{1,0,0}(s,x,y)=&s^2\(1+\(xy+\frac{x}{y}+\frac{y}{x}+\frac{1}{xy}\)s^2 +O(s^4)\) \nn \\
&-s^6\(1+\(xy+\frac{x}{y}+\frac{y}{x}+\frac{1}{xy}\)s^2 +O(s^4)\) \nn \\
=&s^2 P(s,x,y)-s^6 P(s,x,y) \nn \\
=&s^2(1-s^4)P(s,x,y).
\end{align}

For irreps created by operators with more fields, we have to take into account the Bose symmetry between the $\phi$'s. For this purpose, we can use the following generating functional 

\[X(z,s,x,y)=e^{ \sum_{n=1}^{\infty} \frac{z^n}{n} \x_{1,0,0}(s^n,x^n,y^n) }=\sum_{n=1}^{\infty} z^n X_n(s,x,y).\] 
Expanding around $z=0$, the coefficient of $z^n$ is the character for the irrep created by operators with $n$ $\phi$'s. Since the characters of a semi-simple group form a complete basis, we can expand the above character as

\begin{align}
X_n(s,x,y)&=\sum_{\D,j_1,j_2} c_{\D,j_1,j_2} \x_{\D,j_1,j_2}(s,x,y) \\
&=\sum_{\D,j_1,j_2} c_{\D,j_1,j_2}s^{2\D} \x_{j_1}(x)\x_{j_2}(y)P(s,x,y),
\end{align}
where $c_{\D,j_1,j_2}$ is the number that the $(\D,j_1,j_2)$-irrep appears. Using the orthogonality of the $SU(2)$ characters we can write 

\[\sum_{\D,j_1,j_2} c_{\D,j_1,j_2}s^{2\D}=\oint \frac{dx}{2\pi i} \frac{1-x}{x} \oint \frac{dy}{2\pi i} \frac{1-y}{y}\x_{j_1}(x)\x_{j_2}(y) \frac{X_n(s,x,y)}{P(s,x,y)}.\]
Using the expression for $X_n(s,x,y)$ and $P(s,x,y)$ one can determine the coefficients $c_{\D,j_1,j_2}$.

\section{OPE coefficients \label{sec:OPE}}

In this appendix we will determine the OPE coefficients in (\ref{int OPE}). We begin by considering the most general contribution of $V_{1,0}$ in the OPE $V_{n-1,0}(x)V_{n,l}(0)$,

\[ f\sum_{m=0}^l \frac{(u\cdot \hat{x})^{l-m}}{|x|^{\D_{n-1,0}+\D_{n,l}-\D_{1,0}}}|x|^m \(t_m +r_mx\cdot \p +q_m x^\m x^\n\p_\m\p_\n+p_mx^2 \Box+\dots \)(u\cdot \p)^m V_{1,0}(0) \nn.\]
Keeping up to the quadratic terms in $x$, we have

\[V_{n-1,0}(x)V_{n,l}(0) \supset f\frac{(u\cdot \hat{x})^{l}}{|x|^{\D_{n-1,0}+\D_{n,l}-\D_{1,0}}} C(x,\p)V_{1,0}(0),\]
with 

\begin{align}C(x,\p)=&1+r_0 x^\m \p_\m +q_0 x^\m x^\n\p_\m\p_\n+p_0 x^2 \Box+t_1|x| \frac{u\cdot\p}{u\cdot\hat{x}} \nn \\
&+r_1 |x|^2 \frac{\hat{x}\cdot \p \;u\cdot\p}{u\cdot\hat{x}}+t_2 |x|^2 \frac{(u\cdot\p)^2}{(u\cdot\hat{x})^2}+\dots . \label{C with J}
\end{align}
Here we have absorbed $t_0$ into the overall coefficient $f$, so the first term is one. The derivatives are with respect to the argument of $V_{1,0}(y)$, which is set to zero afterwards. This form implies that the $x^2$ contribution to the three-point function with $V_{1,0}(z)$ is 

\begin{align}
\<V_{n-1,0}(x)V_{n,l}(0)V_{1,0}(z)\> \supset& \frac{(u\cdot \hat{x})^{l}}{|x|^{\D_{n-1,0}+\D_{n,l}-\D_{1,0}}}f_0\Big\{1-2\D_{1,0}\(1 -(2\D_{1,0}+2)(\hat{x}\cdot \hat{z})^2 \)q_0\;\|\frac{x}{z}\|^2  \nn \\
&-2\D_{1,0}\(D-(2\D_{1,0}+2)\)\; p_0\; \|\frac{x}{z}\|^2\nn  \\
&-2\D_{1,0}\(1-(2\D_{1,0}+2)\frac{u\cdot\hat{z}\;\hat{x}\cdot\hat{z}}{u\cdot\hat{x}}\)\; r_1\; \|\frac{x}{z}\|^2 \nn \\
&+2\D_{1,0}(2\D_{1,0}+2)f_2\frac{(u\cdot \hat{z})^2}{(u\cdot\hat{x})^2} \|\frac{x}{z}\|^2\Big\}\frac{1}{z^{2\D_{1,0}}}. \label{GF exp} 
\end{align}
On the other hand, conformal invariance fixes the three-point function to be \cite{SPIN,Osborn:1993cr}

\[\<V_{n-1,0}(x)V_{n,l}(0)V_{1,0}(z)\>=\l \frac{\(u \cdot Y\)^l}{|x|^{\D_{n-1,0}+\D_{n,l}-\D_{1,0}-l}\;|z|^{\D_{n,l}+\D_{1,0}-\D_{n-1,0}-l}\;|z-x|^{\D_{n-1,0}+\D_{1,0}-\D_{n,l}+l}} \label{Three point function} \nn,\]
where
\[u\cdot Y=\frac{u\cdot x}{x^2}-\frac{u\cdot z}{z^2}.  \]
Expanding this expression and matching it with (\ref{GF exp}) we can fix the various OPE coefficients. Expanding the numerator, we get 

\begin{align}
(u\cdot Y)^l=& \(\frac{u\cdot\hat{x}}{x}\)^l \(1-\frac{u\cdot \hat{z}}{u\cdot \hat{x}}\;\|\frac{x}{z}\|\)^l   \nn\\
\approx & \(\frac{u\cdot\hat{x}}{x}\)^l  \(1-l\frac{u\cdot \hat{z}}{u\cdot \hat{x}}\;\|\frac{x}{z}\|+\frac{1}{2}l(l-1)\(\frac{u\cdot \hat{z}}{u\cdot \hat{x}}\)^2\;\|\frac{x}{z}\|^2 \),
\end{align}
whereas the denominator becomes

\begin{align}
\frac{1}{|z-x|^{\D_{n-1,0}+\D_{1,0}-\D_{n,l}+l}}&\equiv \frac{1}{|z-x|^{2s}} \nn \\
&=\frac{1}{z^{2s}}\(1-2 \hat{x}\cdot \hat{z}\|\frac{x}{z}\|+\|\frac{x}{z}\|^2\)^{-s}\nn \\
&\approx \frac{1}{z^{2s}}\(1+2s \hat{x}\cdot \hat{z}\|\frac{x}{z}\|-s\|\frac{x}{z}\|^2+2s(s+1)(\hat{x}\cdot \hat{z})^2\|\frac{x}{z}\|^2\), \nn
\end{align}
with $s=\frac{1}{2}(\D_{n-1,0}+\D_{1,0}-\D_{n,l}+l)$. Ultimately, matching the two expressions for this OPE we find

\begin{align}
& \|\frac{x}{z}\|^0:	f=\l		 \nn \\
& \|\frac{x}{z}\|^1:	2\D_{1,0} f r_0=2s	\l	 \nn \\
\frac{(u\cdot \hat{z})}{(u\cdot\hat{x})}& \|\frac{x}{z}\|^1:	2\D_{1,0} f t_1=-l \l	 \nn \\
&\|\frac{x}{z}\|^2:		f\(-2\D_{1,0}q_0-2\D_{1,0}\(D-(2\D_{1,0}+2)\) p_0-2\D_{1,0}r_1\)=- \l s		 \nn \\
(\hat{x}\cdot\hat{z})^2&\|\frac{x}{z}\|^2: 2\D_{1,0}(2\D_{1,0}+2)fq_0=2s(s+1) \l		\nn \\
\frac{u\cdot\hat{z}\;\hat{x}\cdot\hat{z}}{u\cdot\hat{x}}&\|\frac{x}{z}\|^2: 2\D_{1,0}(2\D_{1,0}+2)fr_1=-2s l \l\nn \\
\frac{(u\cdot \hat{z})^2}{(u\cdot\hat{x})^2}&\|\frac{x}{z}\|^2:	2\D_{1,0}(2\D_{1,0}+2)f t_2=\frac{1}{2}l(l-1)\l.
\end{align}
The solution to these equations for $f=\l\neq 0$ is

\begin{align}
& t_2=\frac{l(l-1)}{8\D_{1,0}(\D_{1,0}+1)} \nn \\
& t_1=-\frac{l}{2\D_{1,0}}\nn \\
& r_1= -\frac{(\D_{n-1,0}+\D_{1,0}-\D_{n,l}+l) l}{4\D_{1,0}(\D_{1,0}+1)}\nn\\
& r_0=\frac{\D_{n-1,0}+\D_{1,0}-\D_{n,l}+l}{2\D_{1,0}}\nn\\
& q_0= \frac{(\D_{n-1,0}+\D_{1,0}-\D_{n,l}+l)(\D_{n-1,0}+\D_{1,0}-\D_{n,l}+l+2) }{8\D_{1,0}(\D_{1,0}+1)}\nn\\
& p_0=- \frac{(\D_{n-1,0}+\D_{1,0}-\D_{n,l}+l)(\D_{n,l}-\D_{n-1,0}+\D_{1,0}+l)}{16\D_{1,0}(\D_{1,0}+1)(\D_{1,0}-\d)}. \label{OPE COEFF}
\end{align}

\section{OPE coefficients in $C'(x,\p)$ \label{sec:C'}}

We begin with

\[V_{1,0}(x) V_{2,l}(0) \supset f'\frac{(u\cdot x)^l}{|x|^{\D'_{n,l}+l}} C(x,\p)V_{1,0}(0) ,\]
where $C(x,\p)$ is given by (\ref{C with J}). After applying the operator $\p \cdot D$ on $V_{2,l}$ we will get an expression like

\[V_{1,0}(x) K_{l-1}(0) \supset f'\frac{(u\cdot x)^{l-1}}{|x|^{\D'_{l}+l}} C'(x,\p)V_{1,0}(0) ,\]
where  $\p \cdot D V_{2,l}(0)=\a K_{l-1} $ and 

\[C'(x,\p)=t'_0+p'_0 x^2 \Box+ \dots.\]
For our purposes, it is enough to focus only on the box term, since it contains all the information about the anomalous dimension. At the end of the day one finds

%The first is clear since it already has a box. The second gives a box 

%\[ \p \cdot D \frac{(u\cdot x)^{l-1}}{|x|^{\D'_{n-1,0}+\D'_{n,l}-\D'_{1,0}+l-1}} x\cdot \p \;u\cdot\p =(\frac{d}{2}+l-2) \frac{(u\cdot x)^{l-1}}{|x|^{\D'_{n-1,0}+\D'_{n,l}-\D'_{1,0}+l-1}} \Box +\dots\]

\[t'_0=-\frac{1}{2}l(l+d-3)(d-2+l-\D_l),\]
and
\[p'_0=-\frac{1}{2}l(l+d-3)(d+l-\D_l)p_0 +(\frac{d}{2}+l-2)t_1-(\D_l+l-4)t_2.\]

\section{Three-point function $\<T V_{\D,l} \phi\>$ \label{sec:T ope}}

Following the embedding space notation in \cite{SPIN}, the three-point function of a generic primary operator $V_{\D,l}$ with the energy momentum tensor and a scalar is 

\begin{align}
\< T(P_2) V_{\D,l} (P_3) \Phi (P_1)\>= \frac{1}{(P_1 \cdot P_2)^{\frac{\t_1+\t_2-\t_3+4-2l)}{2}} (P_1 \cdot P_3)^{\frac{\t_1+\t_3-\t_2+2l-4}{2}} (P_2 \cdot P_3)^{\frac{\t_2+\t_3-\t_1+4+2l}{2}}} \nn \\
\scriptstyle{ \times \( a \(\frac{P_1 \cdot C_2 \cdot P_3}{P_1\cdot P_3}\)^2 \(\frac{P_1 \cdot C_3 \cdot P_2}{P_1\cdot P_2}\)^l + b \(\frac{P_1 \cdot C_2 \cdot P_3}{P_1\cdot P_3}\) \(\frac{P_1 \cdot C_3 \cdot P_2}{P_1\cdot P_2}\)^{l-1} C_2 \cdot C_3 + c\(\frac{P_1 \cdot C_3 \cdot P_2}{P_1\cdot P_2}\)^{l-2} (C_2 \cdot C_3)^2 \)} \nn ,
\end{align} 
where $C_{iAB}=Z_{iA} P_{iB} -P_{iA} Z_{iB}$ and $\t_2=d-2$. These three terms correspond to the three conformaly invariant structures we can build. Moreover, using the conservation equation in the embedding space 

\[ \p_2 \cdot D_2 \< T(P_2) V_{\D,l} (P_3) \Phi (P_1)\>=0,\]
we can fix two of the three constants $a,b$ and $c$. The result is

\[a=c  \frac{1}{2}\frac{(d-2) (d + \t_1 - \t_3)}{(d-1)(\t_1+\t_2)-dl} ,\]

\[b=c\frac{2 (-2 + d + \t_1 - \t_3) (\t_1 - \t_3 + d (l - \t_1 + \t_3))}{(d-2)l(l-1)+(d-1)(\t_1-\t_3)(\t_1-\t_3-2l)} .\]   
We can go down to the d-dimensional space using

\[Z_i \cdot Z_j =z_i \cdot z_j \;, \quad P_i\cdot P_j= -\frac{1}{2} x_{ij} ^2 \;, \quad P_i \cdot Z_j = z_j \cdot x_{ij},\]
and the answer reads

\begin{align}
\< T(x) O_{\D,l} (0) \Phi (z)\>= & \quad a \frac{\( \frac{u_2 \cdot (z-x)}{(z-x)^2}+\frac{u_2 \cdot x}{x^2} \)^2 \( \frac{u_3 \cdot z}{z^2}-\frac{u_3 \cdot x}{x^2} \)^l}{z^{\t_1+\t_3-\t_2}\ \; |z-x|^{\t_1+\t_2-\t_3} \;x^{\t_2+\t_3-\t_1}} \nn \\
& -b \frac{\( \frac{u_2 \cdot (z-x)}{(z-x)^2}+\frac{u_2 \cdot x}{x^2} \) \( \frac{u_3 \cdot z}{z^2}-\frac{u_3 \cdot x}{x^2} \)^{l-1} u_2 \cdot I(x) \cdot u_3}{z^{\t_1+\t_3-\t_2}\ \; |z-x|^{\t_1+\t_2-\t_3} \;x^{\t_2+\t_3-\t_1+2}} \nn \\
&+c \frac{ \( \frac{u_3 \cdot z}{z^2}-\frac{u_3 \cdot x}{x^2} \)^{l-2} (u_2 \cdot I(x) \cdot u_3)^2}{z^{\t_1+\t_3-\t_2+4}\ \; |z-x|^{\t_1+\t_2-\t_3-4} \;x^{\t_2+\t_3-\t_1+4}},
\end{align}
where $I_{\m\n}(x)=\d_{\m\n} -2 \frac{x_\m x_\n}{x^2}$.
For simplicity, we will set $u_2=u_3=u$ and then the OPE will take the following form

\[ T(x) O_{\D,l} (0) \supset f \frac{(u \cdot \hat{x})^{l+2}}{|x|^{d-2+\t_3+l+2-\D_1}} (1+q_0 x^\m x^\n\p_\m\p_\n+p_0 x^2 \Box +r_1 |x|^2 \frac{\hat{x}\cdot \p \;u\cdot\p}{u\cdot\hat{x}})\phi(0) \nn .\]
We can proceed as in appendix \ref{sec:OPE} and determine the OPE coefficients to be 

\begin{align*}
&q_0= \frac{(d - 2) (d - 2 + \t_1 - \t_3) (d + \t_1 - \t_3) (d-2 + 2 l - \t_1 + \t_3) (d + 
   2 l - \t_1 + \t_3)}{8 \t_1 (1 + \t_1) ((d - 2) l (l - 1) + (d - 1) (\t_1 - \t_3) (\t_1 - \t_3 - 2 l))}  \nn \\
&p_0=q_0 \frac{2 (2 l - \t_1 + \t_3) (2l-2 +\t_1 + \t_3) + d^3 -   4 d (l^2 - l + 1) + \t_1 d (\t_1 - 2 d -6) - d \t_3 (\t_3 + 4 l - 2)}{(d - 2)(2 - d +   2 \t_1)(d-2 + 2 l - \t_1 + \t_3) (d + 2 l - \t_1 + \t_3)} \\
&r_1={-\frac{(d-2-\t_1-\t_3)(d-2+2l-\t_1+\t_3) (4 (2 l - \t_1 + \t_3) + (d (-2 + l) - 2 l) (d + 2 l - \t_1 + \t_3))}{4 \t_1 (1 + \t_1) ((d - 2) l (l - 1) + (d - 1) (\t_1 - \t_3) (\t_1 - \t_3 - 2 l))} }\nn .
\end{align*}
We are interested in the case $\t_3=3+\g_{3,l}$ and $\t_1=1+\frac{1}{108}\e^2 $. Taking the $\e \rightarrow 0$ limit we arrive at the simple formula

\[p_0=\frac{27 (l+1)^2 }{(l+3)}  \frac{\g_{3,l}}{\e^2 }.\]

%&p_0=\scriptstyle \frac{(d - 2 + \t_1 - \t_3) (d - 2 + 2 l - \t_1 +  \t_3) (2 (2 l - \t_1 + \t_3) (-2 + 2 l +\t_1 + \t_3) + d^3 -   4 d (l^2 - l + 1) + \t_1 d (\t_1 - 2 d -6) - d \t_3 (\t_3 + 4 l - 2))}{8\t_1 (1 + \t_1) (2 - d +   2 \t_1) ((d - 2) l (l - 1) + (d - 1) (\t_1 - \t_3) (\t_1 - \t_3 - 2 l))}  \nn \\

\bibliographystyle{JHEP}
\bibliography{paper}

\end{document}